\documentclass[prl,twocolumn,showpacs,preprintnumbers,amsmath,amssymb]{revtex4}

\usepackage{graphicx}
\usepackage{dcolumn}
\usepackage{bm}

\begin{document}

\preprint{APS/123-QED}

\title{Light Assisted Collisional Loss in a $^{85/87}$Rb Ultracold Optical Trap}

\author{Anthony R. Gorges, Nicholas S. Bingham, Michael K. DeAngelo, Mathew S. Hamilton, and Jacob L. Roberts}
 
\affiliation{%
Department of Physics, Colorado State University, Fort Collins, CO 80523
}%

\date{\today}

\begin{abstract}
We have studied hetero- and homonuclear excited state/ground state collisions by loading both $^{85}$Rb and $^{87}$Rb into a far off resonant trap (FORT). Because of the relatively weak confinement of the FORT, we expect the hyperfine structure of the different isotopes to play a crucial role in the collision rates. This dependence on hyperfine structure allows us to measure collisions associated with long range interatomic potentials of different structure: such as long and short ranged; or such as purely attractive, purely repulsive, or mixed attractive and repulsive. We observe significantly different loss rates for different excited state potentials. Additionally, we observe that some collisional channels' loss rates are saturated at our operating intensities (~15 mW/cm$^{2}$). These losses are important limitations in loading dual isotope optical traps. 
\end{abstract}

\maketitle

Atomic collisions in an ultracold gas in the presence of near-resonant laser light have been studied both experimentally and theoretically since the advent of laser cooling, with both hetero- and homonuclear collisions having been studied \cite{Telles2001,Weiner1999,P1988,Marcassa1993,Sesko1989,Peters1994,Williamson1995,Kawanaka1993,Ritchie1995,Holmes2004,Wang1998,Mancini2004}. These light assisted collisions are responsible for limiting the densities of atoms in magneto-optical traps (MOT) \cite{Wallace1992,Feng1993,Lett1993} and play an important role in limiting the number of atoms that can be loaded into far off-resonant traps (FORT)\cite{Kuppens2000,Ohara2001,Wu2006,Miller1993}. In this work, we describe measurements of light assisted collisions in an ultracold gas composed of a mixture of both $^{85}$Rb and $^{87}$Rb in an optical trap. By measuring the associated loss rates, we can probe the collisions associated with heteronuclear and with homonuclear long range potentials\cite{Marimescu1999}. We observe that the resulting collision rate is a strong function of the excited state potential. Since the two isotopes are so similar, the observed loss rates are directly related to the nature of the excited state potentials, allowing us to probe the dependence of the light assisted collision rate as a function of interatomic potential between atoms while keeping their mass, light scattering, and optical trap temperature and depth characteristics the same. While many of these potentials are complex, some are simple with only purely attractive or purely repulsive characteristics, allowing for an easier interpretation of the observed collision physics. In addition to providing insight into light-assisted collision physics, understanding the behavior of these collisions is useful in understanding the loading dynamics of heteronuclear optical traps, especially those involving two isotopes of the same atom. 

While there has been significant recent activity in the related area of homonuclear and heteronuclear photoassociation\cite{Jones2006}, in photoassociation the relevant internuclear separations are relatively short. For the long ranged potentials studied here, photoassociation will not contribute to the overall loss. Instead a related process, light assisted collisions, is the primary loss mechanism in the overall loading dynamics. Light assisted collisions occur when an atom pair, typically in a ground state, is excited to an excited state interatomic potential (Fig.\ 1 \& 2). After being excited, the atom pair is accelerated along the potential curve until after an excited state lifetime it emits a photon and falls back into the ground state. The photon emitted is less energetic than the one absorbed, and the difference is converted into kinetic energy. If enough kinetic energy is given to the atoms, they can then leave the trap, resulting in loss. To date, the majority of the relevant experimental and theoretical work has been done on ultracold atoms confined to a MOT\cite{Weiner1999}. Compared to a MOT, the light assisted loss is exacerbated in a FORT where trap depths are around 100 $\mu$K compared to the typical 1 K trap depth of a MOT. Because of the shallower depth, loss inducing collisions are much more likely to occur at a longer internuclear separation. Since the difference in hyperfine energies between the two isotopes is much greater than the energy shift due to the interatomic potential at the internuclear radii relevant for loss in the optical trap, the $^{85}$Rb and $^{87}$Rb mixture behaves as a heteronuclear mixture. 

To understand the collision rates due to light assisted collisions as a function of potential, the long range interaction potentials for different combinations of colliding pair excited and ground states were calculated\cite{Marimescu1995}. To calculate these potentials, it was assumed that the interatomic distance between the two atoms was large enough so that exchange interactions could be ignored. Including the hyperfine structure, the dipole-dipole interactions were calculated. The large number of hyperfine and magnetic sublevel combinations give rise to numerous individual interatomic potentials. Fig.\ 1 shows that there are many different types of potentials for heteronuclear collisions: purely repulsive (Fig.\ 1(a)), purely attractive (Fig.\ 1 (b)), or a complex mixture of the two (Fig.\ 1(c-d)). For transitions with mixed potentials, there are numerous avoided crossings and so some initially attractive potentials become repulsive and vice versa. Likewise, the excited state potentials for homonuclear collisions were calculated in the same manner (Fig.\ 2). Unlike heteronuclear collisions, there are no purely attractive or repulsive excited state potentials in homonuclear collisions. The isotopic difference in hyperfine structure produces different homonuclear excited potentials for $^{85}$Rb and $^{87}$Rb, but note that at the highest energy levels the structure of the potentials are qualitatively the same. In addition, homonuclear collisions are longer ranged than heteronuclear collisions. We can choose which individual potential is excited in our experiments, and can thus systematically study loss rates associated with each of the potentials shown in Fig.\ 1 \& 2.  

\begin{figure}
\includegraphics{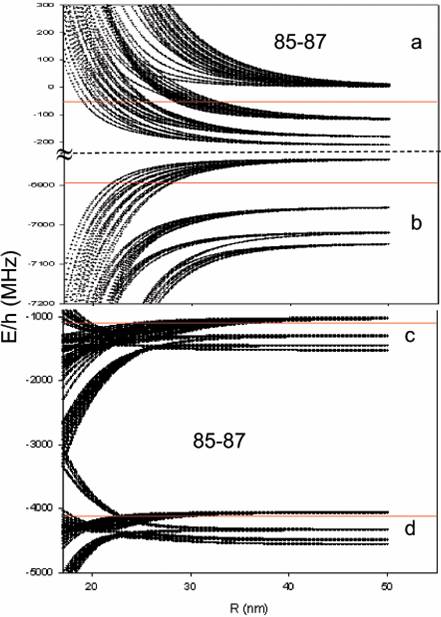}
\caption{\label{label} Excited state potentials for heteronuclear light assisted loss. The zero of the scale is arbitrary. Transitions are accessed from initial ground hyperfine states of: (a) $^{85}$Rb F=3, $^{87}$Rb F=2, (b) $^{85}$Rb F=3, $^{87}$Rb F=1, (c) $^{85}$Rb F=3, $^{87}$Rb F=2, and (d) $^{85}$Rb F=2, $^{87}$Rb F=2. Both (a) and (b) are accessed with a flash light which is detuned by 60 MHz to the red of the $^{85}$Rb cycling transition, where as (c) and (d) are accessed with a flash light which is detuned by 72 MHz to the red of the $^{87}$Rb cycling transition. The horizontal red lines depict laser frequency used to access each transition.}
\end{figure}

To measure these light assisted collisions, we loaded a FORT with either one or simultaneously both isotopes of Rb. Simultaneous loading was accomplished by first capturing and cooling ultracold gases of $^{85}$Rb and $^{87}$Rb into their own MOTs\cite{Suptitz1994}. The MOTs' cooling and hyperfine repump lasers \cite {hyperfine} for the two isotopes were aligned so that the two MOTs overlap in space. Then a 30 W CO$_{2}$ beam was overlapped with the MOTs, and the FORT was loaded by manipulating the MOT laser detuning and hyperfine pump power\cite{Kuppens2000}. The FORT had a trap depth of 120 $\mu$K with trapping frequencies of 450 Hz radial by 35 Hz axial with a typical gas cloud temperature of 15$\mu$K. Standard detunings during the last stage of loading the optical trap for the MOT cooling lasers were 72 MHz and 60 MHz to the red of the cycling transition for $^{87}$Rb and $^{85}$Rb, respectively. Turning the FORT light on and off was performed using an acousto-optical modulator (AOM). After the atoms were loaded into the FORT, all other light (MOT and repump lasers) was shut off and the atoms were held for 100 ms in the FORT to allow for equilibration. Imaging was accomplished through standard absorptive imaging techniques. With our parameters 3.5 million $^{87}$Rb atoms or 4.5 million $^{85}$Rb atoms could be loaded individually. However, when simultaneously loaded the number dropped to around 2 million for each isotope. This reduction is indicative of cross-species light assisted collisions. 

Once the atoms were prepared in the FORT, we illuminated them with a pulse (``flash'') of laser light to induce light assisted collision loss. To drive light assisted collision losses, one of the MOT cooling lasers at its standard detuning was used to couple atom pairs from the ground state to a selected excited state potential ($^{85}$Rb flash intensity was 15 mW/cm$^{2}$ and $^{87}$Rb flash intensity was 25 mW/cm$^{2}$). Typical flash time was 4 ms, but data extending over a range of flash times from 0.5 ms to 20 ms were examined. The main trapping and repump MOT lasers made up the flashing lasers, thus creating an optical molasses. While the complicated polarization structure of an optical molasses is undesirable for these measurements, using a single beam of comparable intensity would produce too much recoil heating to make effective measurements.  We found that there was an elevated initial loss associated with the first few hundred $\mu$s of the flash, while the atoms were being hyperfine pumped.  In order to avoid these complications, we used a 0.5ms flash to establish a baseline and then used longer flashes to measure the loss from that point.  

An additional complication from flashing the atom cloud with an optical molasses came in the form of ``mechanical heating'' of the cloud. The high density of the atoms in the upper hyperfine ground state in the trap can lead to a significant heating of the gas due to rescattering effects\cite{Townsend1995,Walker1990}, depending on the detuning of the pulse.  This could lead to density dependent losses due to subsequent evaporative cooling from the optical trap if the atoms are held there long enough, mimicking a light assisted collision loss. One way this potential systematic uncertainty was mitigated through our choice of flash laser detuning. We also used a two image subtraction technique to measure atom loss that involved holding the atoms in the optical trap for only a short ($\sim$5 ms) time compared to the elastic scattering time. In this technique, the first image was taken while the FORT was held on (in-trap). The second image was taken after the FORT was turned off (out-of-trap) after a 5 ms free expansion time. The in-trap atom count, excluding the FORT region, was then subtracted from the out-of-trap image and this properly accounted for the atoms that had remained in the FORT after the flash, without having to wait until the atoms lost from the FORT had completely fallen away from the imaging region.

\begin{figure}
\includegraphics{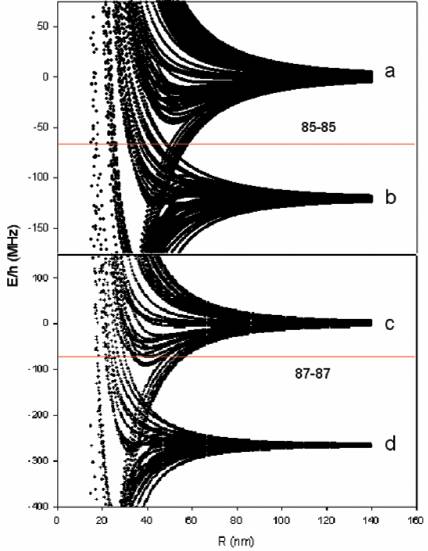}
\caption{\label{label} Excited state potentials for homonuclear light assisted loss. The zero of the scale is arbitrary. Transitions are accessed from initial ground hyperfine states of:(a) and (b) $^{85}$Rb F=3, (c) and (d) $^{87}$Rb F=2. Both (a) and (b) are accessed with a flash light which is detuned by 60 MHz to the red of the $^{85}$Rb cycling transition, where as (c) and (d) are accessed with a flash light which is detuned by 72 MHz to the red of the $^{87}$Rb cycling transition. In contrast to Fig.\ 1 the lettering scheme in this plot refers to specific hyperfine states rather than hyperfine manifolds. The horizontal red lines depict laser frequency used to access each transition.}
\end{figure}

In order to confirm that we were observing density dependent losses, we took data with a single isotope ($^{85}$Rb) that examined the number remaining in the trap as a function of flash time (Fig.\ 3). A one body loss process would appear as a straight line in Fig.\ 3, and since our data do not follow a straight line we confirmed that we were measuring density dependent losses. The number remaining as a function of flash time combined with the measured density of the atoms in the FORT allow for the two-body loss rate (K$_{2}$) to be extracted. Equation (1) and (2) define the differential equations for our measured loss rates.
\vspace{-.05 in}
\begin{eqnarray}
\int{d^3x\;\frac{dn_{85}^{F}}{dt}}=-K_{2(85-85)}^{Fi}\int{d^3x\;(n_{85}^{F})^{2}}\nonumber\\
-K_{2(85-87)}^{FF'i}\int{d^3x\;n_{85}^{F}n_{87}^{F'}}
\end{eqnarray}
\vspace{-.3 in}
\begin{eqnarray}
\int{d^3x\;\frac{dn_{87}^{F'}}{dt}}=-K_{2(87-87)}^{F'i}\int{d^3x\;(n_{87}^{F'})^{2}}\nonumber\\
-K_{2(85-87)}^{FF'i}\int{d^3x\;n_{85}^{F}n_{87}^{F'}}
\end{eqnarray}
Where K$_{2(85-85)}^{Fi}$, K$_{2(87-87)}^{F'i}$, and K$_{2(85-87)}^{FF'i}$ are the light assisted collisional loss rates for homonuclear 85/85, 87/87, and the interspecies 85/87 respectively; $i$ is a label for flash light frequency used. n$_{85}$, n$_{87}$ are the 85 and 87 densities. F, F' are the ground hyperfine states involved in the collision for 85 and 87 respectively.

\begin{figure}
\includegraphics{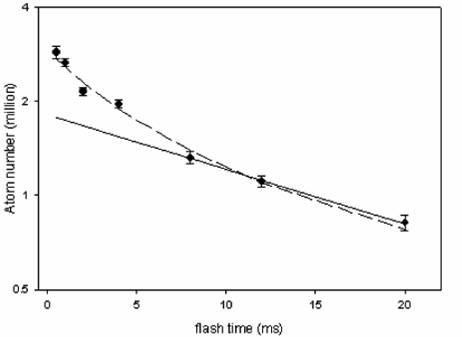}
\caption{\label{label} Log of $^{85}$Rb atom number vs flash time. The dashed line is a fit to the curve assuming two body loss while the straight line is a fit to the last three points. There is a clear deviation of the number evolution from a straight line fit to the last data points.}
\end{figure}

We examined all the transitions which could be resonantly excited under our experimental conditions, except mixed homonuclear ground state distributions. All the measured K$_{2}$ values are reported in Table 1. In addition to the statistical uncertainties shown in the table, there is an additional overall uncertainty of 40$\%$ in the absolute values of the rates due to uncertainy in our density calibration. While two of the measured rates are consistent with zero, the rest of the rates are the same order of magnitude but are different by factors of $\sim$2. These measured rates are much higher than those measured in MOT light assisted collisions for comparable laser intensities and that is expected given the shallow nature of the trap.

The most straightforward collision rates to compare at the qualitative level are those associated with the potentials shown in Fig.\ 1(a-b). Unlike the other cases, the long range excited state potentials are purely repulsive (1a) or attractive (1b). In previous experiments with photoassociation and light assisted collisions, repulsive potentials were used with ``optical shielding'' to reduce collision rates\cite{Sanchez1995,Hoffmann1996,Muniz1997,Walhout1995,Napolitano1997,Shaffer1999}. However, those experiments relied on a resonant excitation with the shielding light where the pair could only gain a maximum kinetic energy which was less than the trap depth, and thus not lost from the trap. For our parameters, a resonant excitation can impart $\sim$10 times the trap depth in kinetic energy; in previous experiments these conditions led to additional loss\cite{Bali1994,Hoffmann1994}. Therefore, if the collision rate is controlled by resonant excitations the loss rate for purely repulsive and attractive potentials should not be markedly different for our parameters. Our observations, however, show the loss rates for the purely repulsive potentials are significantly lower than for the purely attractive potentials.

This degree of suppression of the loss rate is not expected in a semi-classical model of the collision that takes only the excitation rate to the excited state potentials into account. As an example, we performed a loss rate calculation using the Gallagher-Pritchard (GP) model\cite{Gallagher1989} (even though not all the requisite assumptions apply in our parameter range). In making this calculation, we included only radiative escape losses and determined the survival probability for excitation at a given internuclear radius by explicitly integrating the motion of atom pairs on a representative excited state potential to find the time the pair would require to accelerate to the trap escape velocity.  These GP model calculations did not reproduce our observed loss rates. For the detunings used here the GP model gives a loss rate for the purely repulsive potentials which is an order of magnitude greater than that measured. Additionally, the model predicted the purely repulsive potentials would yield a comperable loss to the attractive potentials. 

A better description of the collision dynamics for these potentials can likely be obtained by using a dressed state picture and examining Landau-Zener (LZ) crossing probabilities.   Briefly, as the atoms approach one another during a collision, they encounter an avoided crossing created at the value of internuclear separation R$_{c}$ (Condon radius) where the light resonantly couples the ground and the excited state.  At this avoided crossing, the atoms can either remain in the ground state or adiabatically transfer to the excited state, which could ultimately result in trap loss. This LZ approach has been shown to accurately reproduce more sophisticated theoretical treatments for both attractive\cite{Suominen1998} and repulsive\cite{Suominen1995} potentials.

According the LZ theory\cite{Zener1932,Landau1932}, the probability for making a diabatic crossing is 

\begin{eqnarray}
\vspace{-.05 in}
P=exp[\frac{-\hbar\Omega^{2}}{2\pi\alpha v}]
\vspace{-.3 in}
\end{eqnarray}
Where $\Omega$=$\Gamma$$\sqrt{(I/I_{sat})/2}$ ($\Gamma$ is the natural linewidth), $\alpha$ is the slope of the potential curve at the Condon radius, and $v$ is the velocity of the atom pair. Fig.\ 4(a-b) shows an example of the LZ crossing for the repulsive (4(a)) and attractive (4(b)) potential curves. By comparing the sequence of adiabatic and diabatic crossing that result in the loss in the attractive and repulsive potential cases, we can formulate an LZ prediction for the ratio of those loss rates and compare them to our measurement.

To estimate the ratio between the attractive and repulsive case we model the numerous atom potentials with just one or two representative potentials. In our calculation of $\Omega$ we average over all possible light polarizations and include a Clebsh-Gordon coefficient based on the asymptotic hyperfine state character of the excited state potential. For our parameters, the value of $P$ at the mean velocity is 0.70 and 0.94 for the outermost and innermost avoided crossings in fig. 4(a) and 0.59 for the avoided crossing shown in fig. 4(b). By tallying all of the possible crossings, determining which crossing sequences produce loss, performing a thermal average over all of the collision energies in the cloud, and using equation (3) to estimate the diabatic crossing probability at each avoided crossing, the ratio of the loss probability in the attractive case to the loss probability in the repulsive case can be calculated.

We find that this ratio of attractive to repulsive loss probability is 1.6.  This ratio was obtained ignoring spontaneous emission at R$_{c}$ and hyperfine changing collisions near R=0.  When these spontaneous emission losses hyperfine changing collision events are estimated and included, the ratio does not change significantly, going to 1.3.  The reason for this insensitivity is that approaching the attractive potential case avoided crossing from R=0 is very similar to approaching the outermost repulsive potential avoided crossing from R=$\infty$, and so increasing the loss at these avoided crossings increases the loss probability for both the attractive and repulsive potential case.

While this ratio of probabilities suggests that the attractive potential should produce a larger loss rate than the repulsive, a factor of 1.6 is inconsistent with our measured rates at the 95$\%$ confidence level\cite{probability}.  There are several factors that could explain this disagreement.  First, the LZ model calculates a probability of loss but in order to produce a loss rate an incoming flux needs to be specified as well.  Based on the fact that R$_{c}$ is similar for both the repulsive and attractive curve cases, from purely geometric considerations the incoming flux should be similar.  The collision rates ultimately should be calculated quantum mechanically, though, and that gives the opportunity for destructive and constructive interferences to arise. For instance, during some collisions in the attractive potential case atom pairs will make multiple transits between R=0 and R=R$_{c}$ as they are reflected at R=0 and at the avoided crossing.  The ultimate outward flux of these oscillating atom pairs depends on acquired phases that are not included in our simple model. Also, for our parameters the approximation of reducing the numerous potential curves to a single potential curve is not severe if only average LZ crossing probabilities are considered.  However, this reduction will remove interference effects arising from multiple crossings\cite{Harshawardhan1997,Rangelov2005,Ivanov2008}. We note, though, that our thermal average and magnetic sublevel distribution would likely wipe out some of these interference effects.  

Beyond these interference considerations, problems with this simple LZ picture can also arise because of the assumption of average polarization.  In reality, the atom magnetic sublevel distributions and the light polarization are not uncoupled, and optical pumping will correlate the atom states and the light polarization.  This can produce different effective values of $\Omega$ for the repulsive and attractive cases; though estimates of the impact due to this optical pumping shouldn't change the ratio by more than 20$\%$. Additionally, central to the LZ assumption is that v is constant during the crossing and that the actual potentials can be modeled by replacing them with the appropriate tangent lines at R$_{c}$.  Given that the potentials for our parameters near R$_{c}$ are not as sharp in an absolute sense as in other experiments\cite{Sanchez1995,Muniz1997,Bali1994,Hoffmann1994}, these assumptions may be more questionable in our work.  Our main conclusion is that even for the potentials with the simplest structure, neither the GP model nor the LZ model reproduce the observed ratio of loss rates between the attractive and repulsive cases.  Thus, the dynamics of the collision appear to depend sensitively on the details of the potentials, even for purely attractive and repuslive potentials.

\begin{table}[t]
\begin{tabular}[t]{|c|c|c|} \hline
\multicolumn{1}{|c|}{Loss table} &
 \multicolumn{1}{c|}{87 F=2} &
 \multicolumn{1}{|c|}{85 F=3} \\
\multicolumn{1}{|c|}{($\cdot$10$^{-10}$cm$^{3}$/sec)} &
 \multicolumn{1}{c|}{72 MHz Red} &
 \multicolumn{1}{|c|}{60 MHz Red} \\
\multicolumn{1}{|c|}{} &
 \multicolumn{1}{c|}{of the 87 cycling} &
 \multicolumn{1}{|c|}{of the 85 cycling} \\ \hline
87 F=2 & 6.92(0.52): 2(c) & 0.48(0.35): 1(a) \\ \hline
87 F=1 & - & 2.36(0.68): 1(b) \\ \hline
85 F=3 & 2.22(0.57): 1(c) & 4.75(0.40): 2(a) \\ \hline
85 F=2 & 0.61(0.99): 1(d) & - \\ \hline
\end{tabular}
\caption{\label{label}Measured K$_{2}$ rates.  The isotope and initial hyperfine ground state of each atom in the collision is specified.  Also, the flash light used to induce the loss is specified as well.  The labels for each measured loss rate refer to the specific excited state potentials shown in Figs. 1-2.  The numbers in parenthesis indicate the statistical uncertainties for each measurement.}
\end{table}

While it is relatively straightforward to make comparisons between Fig.\ 1(a-b) due to the simplicity of the excited state potentials, the other accessible excited states have much more complicated structure. In particular, when mixing both attractive and repulsive potentials many avoided crossings are generated in the potentials themselves, as shown in Fig.\ 5. Thus some potentials which are initially attractive during the collision can become repulsive at short range and vice versa, leading to complex dressed state potential curves. While detailed calculations in this system would be difficult, it is reasonable to expect that the presence of repulsive potential curves could mitigate the loss rate. The repulsive curves can turn colliding atom pairs away from short internuclear radii and slow initially accelerated atoms pairs, reducing the loss rate from what it would otherwise be.

\begin{figure}
\includegraphics{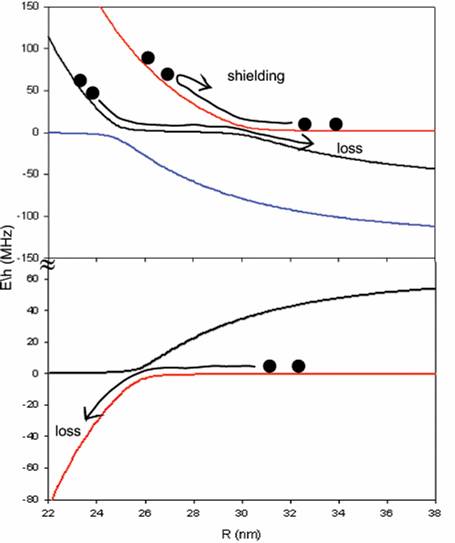}
\caption{\label{label} Representative dressed state potentials used in the LZ calculation of relative loss probabilities.  The zero of the energy scale is selected to correspond to the bare ground state energy at R=$\infty$.  The top figure corresponds to the purely repulsive potential case (corresponding to Fig. 1(a)) and the bottom figure corresponds to the purely attractive potential case (corresponding to Fig. 1(b)).  In the top figure, a sample shielding and sample loss sequence of crossings in indicated.  In the bottom figure, a sample loss sequence is shown.}
\end{figure}

Comparing the loss rates associated with the potentials in Fig.\ 1(c-d) is suggestive of this. The loss rate for the potential associated with Fig.\ 1(d) is less than that for the potential represented in Fig.\ 1(c). One difference between the two potentials is that the one in Fig\. 1(d) is shorter ranged, leading to an expectation of less loss based on the number of atom pairs that collide with sufficiently low impact parameter. In addition, the avoided crossing structure in Fig.\ 1(d) is much sharper, leading to steeper repulsive potential curves, from which a mitigation of the loss rate would be expected.

Fig.\ 2(a-d) show homonuclear excited state potentials for $^{85}$Rb alone (2(a-b)) and $^{87}$Rb alone (2(c-d)). Similar to the heteronuclear potentials shown in Fig.\ 1(c-d), these potentials too have a mixture of attractive and repulsive potentials. However, the homonuclear potentials are longer ranged than the heteronuclear ones, as expected from the 1/R$^{3}$ asymptotic nature of a homonuclear potential as compared to a 1/R$^{6}$ asymptotic nature of a heteronuclear potential. This longer range would suggest, all other things being equal, a larger light assisted collision cross section. Indeed the measured loss rates (see Table 1) indicate that the loss rates are larger for the homonuclear case. Though it is interesting to note that the loss rate for homonuclear $^{85}$Rb collisions is lower than that for homonuclear $^{87}$Rb collisions, despite the Fig.\ 2(a) potential and the Fig.\ 2(c) potential being qualitatively similar and expected to be the potential curves most directly important for the loss. We speculate that this difference is produced from a combination of the difference in the relevant Clebsch-Gordan coefficients for the transitions and the presence of the repulsive potentials in the $^{85}$Rb which extend further out closer to resonance allowing for a higher ``self-shielding'' probability. Once again, the lower loss rate is associated with the potential with the more repulsive character. 

An interesting effect observed in the homonuclear data was that the loss rate was saturated at our intensities\cite{Sukenik1998,Haimberger2006,Kraft2005,Schloder2002,Prodan2003}. To measure saturation, we cut the laser intensity by 1/2 during the flash. The ratio of the 1/2 intensity loss to the full intensity loss is then computed \cite{intensitynote}. Without saturation effects, the loss should scale linearly with the light intensity. With 85 and 87 homonuclear loss, the ratio of the measured K$_{2}$ at half intensity to the rate measured at full intensity is 0.98(12) and 1.08(12) respectively. The fact that no change was observed indicates that the losses are severely saturated at our trap intensities. 

Given the reported results for photoassociation saturation, we first examined whether or not the unitarity limit could be responsible for the observed saturation\cite{Kraft2005,Schloder2002,Prodan2003}. A classical estimation at the most probable collision energy in the cloud indicates that contributions up to h-wave are significant. Including up to h-wave produces a unitarity limit of $\sim$36$\cdot$10$^{-10}$ cm$^3$/sec; much higher than the loss rates measured. This indicates that unitarity is not the cause of the observed saturation. Furthermore, in the unitarity limited regime the scattering rate should not distinguish between $^{85}$Rb and $^{87}$Rb, yet the homonuclear loss rate saturates at different collision rates which suggests the details of the potential must play a role in determining the loss rate. Additionally, the heteronuclear loss rate doesn't appear to saturate and a decrease in the flash intensity by half seems consistent with a decrease in the loss rate by half, as the ratio of the heteronuclear loss of half intensity to full intensity was measured to be 0.60(20).

Rather than unitarity, our results seem consistent with the finite pair formation rate of the atoms in the cloud which is referred to as ``ground state depletion'' in the literature\cite{Gallagher1991}. A classical hard sphere estimate for our experimental conditions shows that for a required close approach internuclear distance of R=72 nm, the maximum pair formation rate is 7$\cdot$10$^{-10}$ cm$^3$/sec; consistent with the observed loss rate. While this is just an estimate, it along with the observed saturation indicates that ground state depletion likely plays a role in these collisions. Since the observed saturation rates are different, the homonuclear potentials must induce some dynamics which alter the collision rates, however. This would be consistent with a model where not every atom pair that collides at the critical radius suggested by ground state depletion is lost.

\begin{figure}
\includegraphics{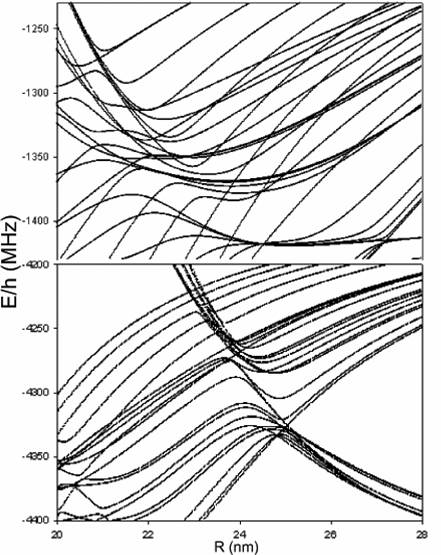}
\caption{\label{label} Excited state potential avoided crossings.  This plot shows two regions of the excited state heteronuclear potentials in more detail to indicate the complicated avoided crossing structure of these potentials.  The top plot is the just the potential shown indicated in fig. 1(c) with its x- and y-axes rescaled.  The bottom plot is the same for the potential indicated in fig. 1(d).}
\end{figure}

There is a concert of rich dynamics and collision physics revealed in the study of light assisted collisional losses. Within the studies presented in this Article, we have measured the hetero- and homonuclear excited state/ground state collision loss rates for $^{85}$Rb and $^{87}$Rb. The measured rates varied significantly depending on the isotopes involved in the collision and their hyperfine state. We observed that a purely repulsive potential reduced the loss rate, in spite of the fact that a direct excitation to the excited state would be expected to produce a large enough gain in kinetic energy to the atom pair to eject them from the FORT. A saturation of the homonuclear collisions at intensities lower than in many other light assisted collision and photoassociation experiments was measured. Estimates suggest that ``ground state depletion'' contributes heavily to the saturation of loss rates. Despite the fact that both $^{85}$Rb and $^{87}$Rb homonuclear loss rates were saturated, they saturate at different rates indicating that the details of the potentials involved in the collision play a significant role in the dynamics. For all measured losses, there is a general trend that the more repulsive the character of the potential (both through purely repulsive states and states that become repulsive via avoided crossings), the lower the loss rate. It is expected that these measurements will be useful in understanding and optimizing the loading and manipulation of multi-isotope traps. 

This work was funded by Air Force Office of Scientific Research, grant number FA9550-06-1-0190.


\begin{thebibliography}{31}
\bibitem{hyperfine}  The 5S$_{1/2}$ F=1$\rightarrow$5P$_{1/2}$ F=2 and 5S$_{1/2}$ F=2$\rightarrow$5P$_{1/2}$ F=3 transitions are used for pumping $^{87}$Rb and $^{85}$Rb respectively.
\bibitem{probability}  Given the size of the uncertainty compared to the measured rate for the 85Rb F=3+87Rb F=2 collision channel, we could not use the usual error propagation techniques that implicitly assume that the uncertainty is small compared to the measured value. Instead, we used Gaussian distributions of the appropriate width centered at the measured central values and then explicitly calculated the probability of various ranges of loss ratios.
\bibitem{intensitynote}  In order to achieve the loading numbers that we did, we needed all the laser power available in the trapping laser. Thus, we couldn't not use the deflected beam from the AOM for the main trapping beam and could not reduce the intensity of the flash continuously to zero without a significant decline in the number of trapped atoms.
\end{thebibliography}
\end{document}